\date{February 17, 1999}

\documentstyle[aps,twocolumn]{revtex}
\begin{document}

\def\no{\noindent}
\def\be{\begin{eqnarray}}
\def\ee{\end{eqnarray}}
\def\non{\nonumber}

\title{BFFT quantization and dynamical solutions of a 
fluid~field~theory}

\author{C. P. Natividade$^{1,2}$\thanks{e-mail: cesar@if.uff.br}
 and H. Boschi-Filho$^3$\thanks{e-mail: boschi@if.ufrj.br}}

\address{$^1$Instituto de F\'\i sica, Universidade Federal Fluminense\\
Avenida Litor\^anea s/n, Boa Viagem, Niter\'oi, 24210-340 Rio de Janeiro, 
Brazil}

\address{$^2$Departamento de F\'\i sica e Qu\'\i mica, 
Universidade Estadual Paulista\\Avenida Ariberto Pereira da Cunha 333, 
Guaratinguet\'a, 12500-000 S\~ao Paulo, Brazil}

\address{$^3$Instituto de F\'\i sica, 
Universidade Federal do Rio de Janeiro\\Caixa Postal 68528, 
Rio de Janeiro, 21945-970 Rio de Janeiro, Brazil}

\maketitle

\begin{abstract}
We study a field theory formulation of a fluid mechanical model.
We implement the Hamiltonian formalism by using the BFFT conjecture 
in order to build a gauge invariant fluid field theory. 
We also generalize previous known classical dynamical field solutions 
for the fluid model.
\end{abstract}




\section{Introduction}
\renewcommand{\theequation}{1.\arabic{equation}}
\setcounter{equation}{0}

The basis of the canonical quantization for systems with infinite 
degrees of freedom has been the powerful character and applicability
of the Dirac method \cite{Dirac}. Despite to current use in different 
systems, alternative formalisms have been developed in order to solve
particular difficulties which come from Dirac's formulation 
\cite{Henneaux}. One of these problems is the role of the first- and
second-class constraints when we identify the classical brackets as 
commutators.
While first-class constraints are related to symmetries the second-class
ones may imply some ambiguities when treated as quantum operators.
The physical status of a theory is chosen by imposing complementary
conditions which are given by the first-class constraints.
In order to avoid the presence of second-class constraints we can separate
it into first-class ones and gauge fixing terms, however there is a special
situation where the constraints are nonlinear so that this procedure fails 
\cite{Banerjee97}. 

An alternative way to circumvent this difficulty is to employ an interesting
machinery proposed by Batalin, Fradkin, Fradkina and Tyutin (BFFT)
\cite{Batalin},
which converts the second-class constraints into first order ones by using
auxiliary fields. Its applicability has been demonstrated in many different
systems involving linear constraints \cite{Fujikawa,Amorim} and also in 
nonlinear cases \cite{Banerjee97,Banerjee94,Rothe98}. 
As we expect, the implementation of the 
above mentioned method through the introduction of new fields gives rise to
a kind of Wess-Zumino terms which turns the resulting effective theory gauge
invariant.

In this paper we discuss the Hamiltonian formalism for a scalar field fluid 
theory from BFFT method point of view. The fluid field theory has been 
introduced as a laboratory to study some classical aspects of membrane 
problem \cite{Hoppe93} but there are also other classical and quantum 
systems which can be described by this model 
\cite{Bazeia&Jackiw,Bazeia,Polychronakos}. 
We can mention, for instance, the hydrodynamical formulation of quantum
mechanics \cite{Madelung} or a dimensional reduction of a relativistic 
scalar field theory \cite{Jevicki}.

Recently, Bazeia and Jackiw \cite {Bazeia&Jackiw} have discussed this 
model by making a careful analysis of the Galileo and Poincar\'e symmetries.
In particular, they obtained dynamical solutions for the original fields
by choosing a singular potential (see also \cite{Bazeia,Polychronakos}). 
In our study we use the BFFT method to
build a gauge invariant theory to obtain the respective generators of the 
extended gauge transformations. As a consequence of this symmetry we show 
that for linear constraints, the Lagrangian is invariant in a similar 
way in respect to that discussed by Amorim and Barcelos \cite{Amorim} 
for chiral bosons theories.

We have organized this paper as follows: In section II we present the fluid
field theory as described by Bazeia and Jackiw. We show that their dynamical
solutions for the singular potential can be generalized to other potentials
leading to diverse physical systems. Section III  is dedicated to 
the explanation of the BFFT method applied to the fluid field theory.
Finally in section IV we present an analysis of the results obtained and
give our conclusions. We have also included two appendices where some 
technical calculations are given.

\section{The Model its Symmetries and Solutions}
\renewcommand{\theequation}{2.\arabic{equation}}
\setcounter{equation}{0}

Let us consider a fluid dynamical model \cite{Landau} 
described by the following Lagrangian 
in $d$ dimensional {\bf r} space, evolving in time $t$:
\begin{equation}\label{L}
L=\int d^dr\; \left( \theta\dot\rho -{1\over 2}\rho\nabla\theta \cdot
\nabla\theta - V(\rho)\right)
\end{equation}

\noindent 
where $\rho = \rho (t,{\bf r})$, $\theta = \theta (t,{\bf r})$ and the
over dot means time differentiation. In a usual fluid mechanical model
$\rho$ is the mass density and $\theta$ is the velocity potential,
${\bf v}=\nabla\theta$.
Then, we have the equations of
motion:
\begin{equation}\label{motion1}
\dot\rho = \nabla \cdot (\rho\nabla\theta)
\end{equation}
\begin{equation}\label{motion2}
\dot\theta=-{1\over 2}(\nabla\theta)^2
- {\delta\over \delta\rho}\int d^dr\; V
\end{equation}

\no These equations of motion 
 are recognized as the conventional ones for isentropic irrotational 
fluids \cite{Landau}. As this system is a nonrelativistic one it 
has naturally the Galilean symmetry \cite{Mukunda}, {\sl i. e.},
it is invariant under the Galilean Group which generators are
\begin{eqnarray}
H&=&\int d^dr\; {\cal E}; 
\qquad\qquad
{\cal E}=\frac 12\rho(\nabla\theta)^2+V(\rho)\label{E}
\\
{\mathbf P}&=&\int d^dr\; {\cal P};
\qquad\qquad
{\cal P}=\rho\nabla\theta
\\
J^{ij}&=&\int d^dr\; {\cal J}^{ij};
\qquad\quad
{\cal J}^{ij}=r^i{\cal P}^j-r^j{\cal P}^i
\\
{\mathbf B}&=&\int d^dr\; {\cal B};
\qquad\qquad
{\cal B}= t {\cal P}-{\bf r} \rho\label{B}
\\
N&=&\int d^dr\; \rho\label{N}
\end{eqnarray}

\noindent which respectively give time and space translation 
and space rotation, in addition to the Galileo boost and 
``charge'' generator $N$ which in this case is the total mass 
of the fluid, being naturally conserved. 
The Poisson brackets of these generators close
under an algebra corresponding to the Galileo group. 
For instance,
\be
\{B^i,P^j\}=\delta^{ij}N.
\ee

 A connection with the membrane problem
\cite{Hoppe93} is done when $d=2$ and
\begin{equation}\
V(\rho)={g\over \rho}\label{V}
\end{equation}

\noindent where $g$ is the coupling constant (this potential also
connects the fluid field theory problem to $d$-branes in $d+1$ space 
dimensions \cite{Bazeia,Polychronakos}). 
In this case, the action 
$I=\int dt L$ is invariant under time rescaling $t\to e^{w} t$ 
generated by (dilation) 
\be
D=
\int d^dr (t{\cal E}-\rho\theta)\label{D}
\ee

Another symmetry of this action is given implicitly by \cite{Bazeia&Jackiw}
\be
t &&\to 
T(t,{\bf r})= t + \bf{w}\cdot{\bf r} + \frac 12 {\bf w}^2\theta(T,R)
\non\\
{\bf r}&&\to
{\bf R}(t,{\bf r})= {\bf r} + {\bf w} \theta(T,R)
\non
\ee
where
\be
\theta (T,R) = \theta (t,{\bf r} - {\bf w} t)
+ {\bf w} \cdot {\bf r}
-\frac 12 {\bf w}^2t
\non
\ee
which are generated by
\be
{\bf G}=
\int d^dr({\bf r}{\cal E}-\theta{\cal P})\,.\label{G}
\ee

\no 
The geometrical meaining of ${\bf G}$ is not clear, as pointed out in Ref. 
\cite{Bazeia&Jackiw}, however one should note that $D$ and ${\bf G}$ 
depend on the velocity potential $\theta$ which is meaningful only in the 
irrotational case. They speculated that, since the fluid field model
corresponds to a gauge-fixed version of the relativistic membrane in the 
light-cone, the symmetry generated by ${\bf G}$ may be a residual gauge 
invariance of that model.
As is well known, the Galileo group (\ref{E})-(\ref{N}) together with the
generators (\ref{D}) and (\ref{G}) defined in (d+1) dimensions is 
isomorphic to a Poincar\'e group defined in (d+1,1) dimensions 
\cite{Susskind68}. 

Bazeia and Jackiw presented some solutions for this system
with the potential (\ref{V}) which are 
\begin{equation}
\theta(t,{\bf r})=-{r^2\over 2(d-1)t}
\end{equation}
\begin{equation}
\rho(t,{\bf r})=\sqrt{2g\over d}(d-1){|t|\over r}
\end{equation}

\noindent valid for $d>1$. They discuss other solutions for the
free case (a particular case of the above solution) and also some
solutions for $d=1$, which are not our main concern here. Other 
solutions in different dimensions can also be found in \cite{Bazeia}.

Here we note that it is possible to extend the above results
considering now the following potential:
\begin{equation}
V(\rho)= {g\over \rho^n}\label{Vn}
\end{equation}

\noindent 
which describe ideal polytropic gases , {\sl i. e.}, a gas in which the 
pressure is proportional to a power of the density \cite{Chandrasekhar}, 
for which we find the solutions
\begin{equation}
\theta(t,{\bf r})= - {r^2\over [d(n+1)-2]t}
\end{equation}
\begin{equation}
\rho(t,{\bf r})=
\left\{ {ng [d(n+1)-2]^2 t^2 \over d(n+1) r^2} \right\}^{1/(n+1)}
\end{equation}

\noindent which are valid in $d>1$ space dimensions, generalizing the 
solutions obtained by Bazeia and Jackiw \cite{Bazeia&Jackiw}. 
Their solution is a particular case of the above class of solutions 
which can be recovered when we take $n=1$ in the above equations.
As we are going to show in the 
following section, the above fluid field theory also admits a gauge 
symmetry which is respected by a general potential $V(\rho)$.


\section{BFFT Quantization and Gauge Symmetry}
\renewcommand{\theequation}{3.\arabic{equation}}
\setcounter{equation}{0}

Let us now construct a gauge invariant version of the model described 
above using the method developed by Batalin, Fradkin, Fradkina, and
Tuyutin (BFFT) \cite{Batalin} which transforms second-class constraints
into first-class ones \cite{Amorim}.

Considering the Lagrangian (\ref{L}) we obtain the primary constraints
\begin{eqnarray}
\chi_\rho&=&\Pi_\rho-\theta\approx 0
\\
\chi_\theta&=&\Pi_\theta\approx 0
\end{eqnarray}

\noindent which satisfy the algebra
\begin{eqnarray}
\left\{\chi_\rho,\chi_\theta\right\}
&=&-\epsilon_{\rho\theta}\ \delta(x-y)
\nonumber\\
&\equiv&\Delta_{\rho\theta}(x,y)\label{Delta}
\end{eqnarray}

\no Then, the primary Hamiltonian is given by
\be
H_p&=&\int d^d r \Big( \Pi_\rho \dot\rho +\Pi_\theta\dot\theta
-L+\lambda_\rho\chi_\rho+\lambda_\theta\chi_\theta \Big)
\non\\
&=&\int d^d r \Big( \left(\Pi_\rho-\theta\right)\dot\rho
+\Pi_\theta\dot\theta
+\frac 12 \rho (\nabla\theta)^2
\non\\
&&\qquad\qquad
		+V(\rho)+\lambda_\rho\chi_\rho+\lambda_\theta\chi_\theta\Big)
\non\\
&=&\int d^d r \Big( \frac 12 \rho (\nabla\theta)^2
+V(\rho)+\tilde\lambda_\rho\chi_\rho+\tilde\lambda_\theta\chi_\theta\Big)
\label{H}
\ee

\no where we defined
$\tilde\lambda_\rho=\lambda_\rho+\dot\rho$, 
$\tilde\lambda_\theta=\lambda_\theta+\dot\theta$. 
The consistency condition for the constraints determine the
fields $\tilde\lambda_\rho$, $\tilde\lambda_\theta$ and there are
no other constraints.

Before we implement the BFFT method it is necessary here 
to make a brief review of it. 
For a more comprehensive and elegant discussion see Refs. 
\cite{Batalin,Amorim}. Let us now begin by extending 
the phase space including the new fields $\varphi_\rho$ and 
$\varphi_\theta$ which satisfy the algebra
\be\label{omega}
\left\{\varphi_\rho,\varphi_\theta\right\}=\omega_{\rho\theta}(x,y)
\ee

\no such that the new constraints $\Omega_\rho$, $\Omega_\theta$
should be of first-class and could be  written in general as
\be \label{Omega}
\Omega_\beta=\chi_\beta+\sigma_{\beta\alpha}\varphi^\alpha
\ee

\no where $\sigma_{\beta\alpha}=\sigma_{\beta\alpha}(\rho,\theta)$.
The central idea of the BFFT method is to write the first-class
constraints in terms of the second-class ones as 
\be\label{Omegan}
\Omega_\beta=\sum_{n=0}^\infty\chi_\beta^{(n)},
\ee

\no with the condition $\chi_\beta^{(0)}\equiv\chi_\beta$. 
So, $\chi_\beta^{(n)}$ is of nth order in the field $\varphi_\alpha$.
The new constraints defined by Eqs. (\ref{Omega})-(\ref{Omegan}) 
satisfy the relation $\{\Omega_\alpha,\Omega_\beta\}=0$ and then
\be\label{BFT1}
&&\{\chi_\alpha,\chi_\beta\}_{(\rho,\theta)}
+\{\chi_\alpha^{(1)},\chi_\beta^{(1)}\}_{(\varphi)}
=0
\\ \label{BFT2}
&&\{\chi_\alpha,\chi_\beta^{(1)}\}_{(\rho,\theta)}
+\{\chi_\alpha^{(1)},\chi_\beta\}_{(\rho,\theta)}
\non\\
&&\quad+\{\chi_\alpha^{(1)},\chi_\beta^{(2)}\}_{(\varphi)}
+\{\chi_\alpha^{(2)},\chi_\beta^{(1)}\}_{(\varphi)}=0
\\
&&\{\chi_\alpha,\chi_\beta^{(2)}\}_{(\rho,\theta)}
+\{\chi_\alpha^{(1)},\chi_\beta^{(1)}\}_{(\rho,\theta)}
\non\\
&&\quad+\{\chi_\alpha^{(2)},\chi_\beta\}_{(\varphi)}
+\{\chi_\alpha^{(1)},\chi_\beta^{(3)}\}_{(\varphi)}
\non\\
&&\quad+\{\chi_\alpha^{(2)},\chi_\beta^{(2)}\}_{(\varphi)}
+\{\chi_\alpha^{(3)},\chi_\beta^{(1)}\}_{(\varphi)}
=0.\label{BFT3}
\ee

\no Here, we are using the notation $\{\;,\;\}_{(\rho,\theta)}$, 
$\{\;,\;\}_{(\varphi)}$ referring to the Poisson brackets of the 
pairs $(\rho,\theta)$ and $(\varphi_\rho,\varphi_\theta)$. 
From Eqs. (\ref{Delta}), (\ref{omega}), (\ref{Omega}), (\ref{BFT1}) 
and using that $\{\Omega_\alpha,\Omega_\beta\}=0$ we have
\be
\Delta_{\rho\theta}=-\sigma_{\rho\alpha}\ \omega^{\alpha\beta}
\sigma_{\theta\beta},
\ee

\no which for the fluid field problem can be written as
\be 
\Delta_{\rho\theta}&=&-\ \epsilon_{\rho\theta}\ \delta(x-y)
\non\\
&=&-\int dz\;dz^\prime\; \sigma_{\rho\alpha}(x,z)\ 
\omega^{\alpha\beta}(z,z^\prime)\ 
\sigma_{\theta\beta}(y,z^\prime).
\ee

\no As $\omega^{\alpha\beta}$ is obtained from second-class constraints
$\varphi_\alpha$ we can choose 
$\omega^{\alpha\beta} = \epsilon_{\alpha\beta}\ \delta(x-y)$ which
implies $\sigma_{\rho\alpha} = \epsilon_{\rho\alpha}\ \delta(z-x)$ so that
\be
\Omega_\rho(x)&=&\chi_\rho(x)+\int dz\ \delta(z-x)\ \varphi_\theta(z)
\non\\
&=& \chi_\rho(x)+\sigma_{\rho\theta}\ \varphi^\theta(x)
\ee
and similarly
\be
\Omega_\theta(x)&=&\chi_\theta(x)+\sigma_{\theta\rho}\ \varphi^\rho(x)
\ee

\no The next step is to include corrections to the canonical Hamiltonian.
We remark that in this formalism any dynamical function $A(\rho,\theta)$
can also be properly modified in order to be strong involutive with 
first-order constraints. So, if $\tilde{A}(\rho,\theta,\varphi)$ is this 
quantity we have
\be 
\{\Omega_\rho,\tilde A\}=0
\ee
with the boundary condition
\be
\tilde A(\rho,\theta,0)= A(\rho,\theta)\,.
\ee

In order to generate $\tilde{A}$ we can repeat the same steps for the 
obtainment of $\Omega_\rho$ above, {\sl i. e.} we consider the expansion
\be
\tilde{A}=\sum_{n=0}^\infty A^{(n)}\,,
\ee
where $A^{(n)}$ is a term of order $n$ in the field $\varphi$. 
Consequently,from Eqs. (\ref{BFT1})-(\ref{BFT3}) rewritten for 
$A^{(n)}$ and the condition $A^{(0)}=A$, we have
\be 
A^{(1)}=-\varphi^\alpha\omega_{\alpha\beta}\,\sigma^{\beta\gamma}
\{\chi_\gamma,A\}\,,
\ee

\no where $\omega_{\alpha\beta}=(\omega^{\alpha\beta})^{-1}$
and $\sigma^{\beta\gamma}=(\sigma_{\beta\gamma})^{-1}$.
An equation analogous to (\ref{BFT2}) for $A^{(2)}$ gives
\be 
\{\chi^{(1)}_\rho,A^{(2)}\}=-G^{(1)}_\rho
\ee

\no such that
\be 
G^{(1)}_\rho &=& \{\chi_\rho,A^{(1)}\}_{(\rho,\theta)}
+\{\chi_\rho^{(1)},A\}_{(\rho,\theta)}
\non\\
&&\quad+\{\chi_\rho^{(2)},A^{(1)}\}_{(\varphi)}.
\ee

\no Then, we have for $A^{(2)}$
\be 
A^{(2)}=-\frac 12 \varphi^\alpha\omega_{\alpha\beta}\sigma^{\beta\gamma}
G^{(1)}_\gamma.
\ee

\no and in general for $n\ge 1$
\be \label{An+1}
A^{(n+1)}=-\frac 1{n+1}\varphi^\alpha\omega_{\alpha\beta}\,
\sigma^{\beta\gamma}
G^{(n)}_\gamma\;,
\ee

\no with the auxiliary condition $G^{(0)}_\rho=\{\chi_\rho,A\}$ so that
\be 
G^{(n)}_\rho &=&
\sum_{m=0}^n\{\chi_\rho^{(n+m)},A^{(m)}\}_{(\rho,\theta)}
\non\\
&&+\sum_{m=0}^{n-2}\{\chi_\rho^{(n-m)},A^{(m+2)}\}_{(\varphi)}
\non\\
&&+\{\chi_\rho^{(n+1)},A^{(1)}\}_{(\varphi)}.
\ee


\subsection{\it Particular Case: $\sigma_{\alpha\beta}$ independent of 
$(\rho,\theta)$}

In the particular case where $\sigma_{\alpha\beta}$ does not depend on
$(\rho,\theta)$ the $A^{(n)}$ are still given by Eq.  (\ref{An+1}) but
\be 
G^{(n)}_\rho=\{\chi_\rho,A^{(n)}\}.
\ee

\no Let us analyze this particular case further with the additional 
hypothesis that the second-class constraints are all linear.
Then, we can write
\be 
A^{(n+1)}
&=&-\frac 1{n+1}\varphi^\alpha\omega_{\alpha\beta}\sigma^{\beta\gamma}
\{\chi_\gamma,A^{(n)}\}
\non\\
&=&-\frac 1{n+1}\varphi^\alpha\omega_{\alpha\beta}\sigma^{\beta\gamma}
\{\chi_\gamma,z^i\}\frac{\partial}{\partial z^i} A^{(n)}
\non\\
&=&-\frac 1{n+1}\varphi^\alpha k^i_\alpha \partial_i A^{(n)}
\label{Aki}
\ee

\no where we have used the Jacobi identity and the defined 
$z^i$ is a generalized phase space coordinate, in the sense
that it could be either a canonical coordinate or a canonical momentum.
Note that 
$k^i_\alpha\equiv \omega_{\alpha\beta}\sigma^{\beta\gamma}
\{\chi_\gamma,z^i\}$
 is a constant matrix since the constraints are linear,
$\{\chi_\gamma,z^i\}=constant$ and $\sigma^{\beta\gamma}$
is a constant matrix too in this particular case.
Using Eq. (\ref{Aki}) iteratively one finds
\be 
A^{(n)}
=\frac {(-1)^n}{n!}(\varphi^\alpha k^i_\alpha \partial_i)^n A
\ee

\no so that
\be
\tilde A(\rho,\theta;\varphi)
&\equiv&\sum_{n=0}^\infty A^{(n)}
\non\\
&=&\sum_{n=0}^\infty \frac {(-1)^n}{n!}
(\varphi^\alpha k^i_\alpha \partial_i)^n A
\non\\
&=&\exp(\varphi^\alpha k^i_\alpha \partial_i) A
\ee

\no and then in this case the operator $\tilde A$ will be of the form
\be 
\tilde A(z^i,\varphi^\alpha)
=A(z^i-\varphi^\alpha k^i_\alpha) .
\ee


\subsection{\it General Case: $\sigma_{\alpha\beta}$ as as function of
$(\rho,\theta)$}

Let us now return to the discussion of the general case and 
construct the extended Hamiltonian.
Noting that for $n\ge 2$, $\chi_\rho^{(n)}=0$, so that
\be
H_c^{(n+1)}&=&-\frac 1{n+1}\int dx\; dy\; dz \;
\non\\
&&\varphi_\alpha(x)(\omega_{\alpha\beta})^{-1}
(\sigma_{\beta\gamma})^{-1}
G_\gamma^{(n)}\;,\label{Hn+1}
\ee

\no where $G_\gamma^{(n)}$ is given by
\be
G_\gamma^{(n)}=\{\chi_\gamma,H_c^{(n)}\}\label{Gn}
\ee

\no Since $(\omega_{\alpha\beta})^{-1}$ and
$(\sigma_{\beta\gamma})^{-1}$ are proportional to
Dirac delta functions, we have:
\be
H_c^{(0)}=\int dx\; [\frac 12 \rho (\nabla\theta)^2-V(\rho)]
\ee

\no so that
\be
G_\rho^{(0)}&=&\{\chi_\rho,H_c^{(0)}\}
\non\\
&=&-\frac 12 (\nabla\theta)^2+\partial_\rho V
\ee

\no and also
\be
G_\theta^{(0)}&=&\{\chi_\theta,H_c^{(0)}\}
\non\\
&=&-\frac 12 \rho \nabla^2\theta
\ee

\no so that the correction $H_c^{(1)}$ is given by
\be
H_c^{(1)}
&=&-\int dx\, 
\Big\{\Big[\frac 12 (\nabla\theta)^2-\partial_\rho V\Big]\varphi_\rho
\non\\
&&\qquad\qquad-\frac 12 (\rho \nabla^2\theta)\varphi_\theta\Big\}
\ee

\no Continuing the iteration process and summing up all the 
contributions we find that the canonical Hamiltonian is then given by 
(see the Appendix A)
\be
H_c&=&H_c^{(0)}+\int dx \;
\Big[-\lambda_\theta\varphi_\rho +\lambda_\rho\varphi_\theta
-\frac 12 (\partial_\rho^2 V)\varphi_\rho^2
\non\\
&&\qquad\qquad\qquad
+\dots+\frac {(-1)^n}{n!}(\partial_\rho^n V)\varphi_\rho^n
+\dots\Big]
\non\\
&=&H_c^{(0)}+\int dx \;
\Big[-\lambda_\theta\varphi_\rho +\lambda_\rho\varphi_\theta
\non\\
&&\qquad\qquad
+\sum_{n=0}^\infty\frac {(-1)^{n+2}}{(n+2)!}
(\partial_\rho^{(n+2)} V)\varphi_\rho^{(n+2)}\Big]
,\label{Hc}
\ee

\no where the term corresponding to $H_c^{(1)}$ is contained in
$\lambda_\theta=\frac 12 (\nabla\theta)^2-\partial_\rho V$.
Note that from the definition (\ref{Omega}) the first-class
constraints are given by
\be 
\Omega_\rho&=&\chi_\rho+\sigma_{\rho\theta}\varphi^\theta
\label{Omegarho}
\\
\Omega_\theta&=&\chi_\theta+\sigma_{\theta\rho}\varphi^\rho
\label{Omegatheta}
\ee

\no where $\sigma_{\rho\theta}=-\sigma_{\theta\rho}=1$. 
 Let us now go back to the canonical Hamiltonian 
Eq. (\ref{Hc}) and analyze its last term. We note that
\be
&&\sum_{n=0}^\infty\frac {(-1)^{n+2}}{(n+2)!}
(\partial_\rho^{(n+2)} V)\varphi_\rho^{(n+2)}
\non\\
&&\qquad\qquad
=\sum_{n=0}^\infty \Theta(n-1)\frac {(-1)^{n}}{n!}
(\partial_\rho^{n} V)\varphi_\rho^{n}
\non\\
&&\qquad\qquad
=-V(\rho)+\varphi_\rho\partial_\rho V(\rho)
-e^{-\varphi_\rho\partial_\rho}V(\rho)\;,
\ee

\no where $\Theta(x)$ is the Heavside function.
This way we have 
\be
H_c
&=&\int dx \;
[-\lambda_\theta\varphi_\rho +\lambda_\rho\varphi_\theta
-e^{-\varphi_\rho\partial_\rho}V(\rho)]
\non\\
&=&\int dx\;
[\frac 12 \rho (\nabla\theta)^2
-\frac 12 (\nabla\theta)^2\varphi_\rho
\non\\
&&\qquad\qquad
-\frac 12 \rho\nabla^2\theta\varphi_\theta
-e^{-\varphi_\rho\partial_\rho}V(\rho)].\label{H-c}
\ee

\no In order to find the corresponding Lagrangian, 
we identify $\varphi\equiv\varphi_\rho$ and 
$\Pi_\varphi\equiv\varphi_\theta$ as a pair of canonical 
conjugate coordinates and write the generating functional
\be
{\cal Z}
&=&{\cal N}\int [d\rho][d\Pi_\rho][d\theta][d\Pi_\theta]
[d\varphi][d\Pi_\varphi]
\non\\ 
&&\quad
\times\delta(\Pi_\rho-\theta+\Pi_\varphi)\delta(\Pi_\theta-\varphi)
\non\\
&&\quad
\times\exp\big\{i\int dx \;
\big[\Pi_\rho\dot\rho+\Pi_\theta\dot\theta
+\Pi_\varphi\dot\varphi
\non\\ 
&&\qquad\qquad
-\frac 12 \rho (\nabla\theta)^2
+\frac 12 (\nabla\theta)^2\varphi
\non\\
&&\qquad\qquad
+\frac 12 \rho\nabla^2\theta\Pi_\varphi
+e^{-\varphi\partial_\rho}V(\rho)\big]\big\}
\ee

\no where the delta functions represent the first-class constraints.
Noting that
\be
&&{\rm (i)}\qquad
\int[d\Pi_\theta]\delta(\Pi_\theta-\varphi)
\exp\{i\int dx\; \Pi_\theta\dot\theta\}
\non\\
&&\qquad\qquad=\exp\{i\int dx\; \varphi\dot\theta\}
\non\\
&&{\rm (ii)}\qquad
\int[d\Pi_\rho]\delta(\Pi_\rho-\theta-\Pi_\varphi)
\exp\{i\int dx \;\Pi_\rho\dot\rho\}
\non\\
&&\qquad\qquad=\exp\{i\int dx\; (\Pi_\varphi+\theta)\dot\rho\}
\non
\ee

\no and substituting these results into the generating functional
we find
\be
{\cal Z}
&=&{\cal N}\int [d\rho][d\theta][d\varphi][d\Pi_\varphi]
\non\\
&&\qquad
\exp\big\{i\int dx \;
\big[\theta\dot\rho+\Pi_\theta\dot\theta+\Pi_\varphi\dot\varphi
\non\\
&&\qquad\qquad
-\frac 12 \rho (\nabla\theta)^2
+\frac 12 (\nabla\theta)^2\varphi
\non\\
&&\qquad\qquad
+\frac 12 \rho\nabla^2\theta\Pi_\varphi
+e^{-\varphi\partial_\rho}V(\rho)\big]\big\}.
\ee

\no The functional integral over $\Pi_\varphi$ gives
\be
&&\int [d\Pi_\varphi] \exp\{i\int dx\; 
\Pi_\varphi(\dot\rho+\dot\varphi+\frac 12 \rho\nabla^2\theta)\}
\non\\
&&\qquad\qquad
=\delta(\dot\rho+\dot\varphi+\lambda_\rho)
\non
\ee

\no so that the generating functional reads
\be
{\cal Z}
&=&{\cal N}\int [d\rho][d\theta][d\varphi]
\delta(\dot\rho+\dot\varphi+\lambda_\rho)
\non\\
&&\times\exp\big\{i\int dx \;
\big[\theta\dot\rho+\varphi\dot\theta
-\frac 12 (\rho-\varphi)(\nabla\theta)^2 
\non\\
&&\qquad\qquad
+e^{-\varphi\partial_\rho}V(\rho)\big]\big\}
\ee

\no and then the extended Lagrangian density is given by
\be
\tilde{\cal L}=
\theta\dot\rho+\varphi\dot\theta
-\frac 12 (\rho-\varphi)(\nabla\theta)^2 
+e^{-\varphi\partial_\rho}V(\rho).
\ee

\no In the limit in which the auxiliary field vanishes, 
$\varphi\to 0$, we get back the original Lagrangian (\ref{L})
as it should. Let us now look at the last term of this Lagrangian
which involves the potential $V(\rho)$. Assuming that the potential
function can be expanded in a power series we have
\be
e^{-\varphi\partial_\rho}V(\rho)
=V(\rho-\varphi)
\ee

\no  So, we can apply 
the above equation for a wide range of functions $V(\rho)$, 
as for example, the potentials discussed in section II.

\no Then, the extended Lagrangian density 
 after a partial integration becomes
\be
\tilde{\cal L}=
-(\rho-\varphi)\dot\theta
-\frac 12 (\rho-\varphi)(\nabla\theta)^2 
+V(\rho-\varphi)
\ee

\no so that the Lagrangian is invariant under the exchange
$\rho\to\rho-\varphi$ which is the gauge symmetry of the model.
Since the first-class $\Omega$ is strongly involutive with canonical
Hamiltonian (see the Appendix B) it is easy to check the invariance of
$\tilde{\cal L}(\rho,\theta,\varphi)$.

Now, we can look at the consequences of this gauge symmetry on
the previous known symmetries for the fluid dynamical model.
The Galileo and Poincar\'e groups in the gauged model can be 
obtained from the the density generators of the non-gauge
model, Eqs. (\ref{E})-(\ref{N}), (\ref{D}) and (\ref{G}) 
simply through the shift 
$\tilde{\cal O}=e^{-\varphi\partial_\rho}{\cal O}$,
so that the original Galileo and Poincar\'e invariances 
of the fluid field model are preserved by the introduction of
the auxiliary field which bring to it a gauge symmetry.



\section{Conclusions}

In this article we have studied the fluid field theory \cite{Hoppe93} 
for which we found a class of classical solutions which recover previous
known particular cases \cite{Bazeia&Jackiw}. 

Then, by means of the BFFT 
formalism \cite{Batalin} we extended the original phase space by 
including new fields which permitted the transformation of 
the set of second-class constraints into a first-class one. 

We have analyzed a situation where we found an extended gauge symmetry 
for an arbitrary potential with linear constraints and a kind of a 
Wess-Zumino Lagrangian was built.
As a result we have obtained a new gauge invariant system.
This new system 
may be of interest to the membrane problem related to Lagrangian 
(\ref{L}) since that formulation corresponds to a gauge fixed version 
in the light-cone gauge.

As a final remark it is important to mention that the procedure discussed
here could also be applied successfully 
to a situation where nonlinear constraints
were involved, as is well known in general for the BFFT method.

\acknowledgements
H.B.-F. would like to acknowledge interesting discussions with D. Bazeia, 
R. Jackiw, J. Hoppe and C. Sigaud on the model presented in section II 
and CTP-MIT where part of this work was done. 
C.P.N. was partially supported by CNPq and 
H.B.-F.  by CNPq, FINEP and FUJB 
-- Brazilian research agencies.


\section*{Appendix A}
\renewcommand{\theequation}{A.\arabic{equation}}
\setcounter{equation}{0}

In this Appendix we give some details of the iteration process necessary
to construct the canonical Hamiltonian in the case of the linear constraint
discussed in section III. Using Eqs. (\ref{Hn+1}) and (\ref{Gn}) we found 
the first correction, $H_c^{(1)}$. 
For the next correction we have:
\be
G_\rho^{(1)}&=&\{\chi_\rho,H_c^{(1)}\}
\non\\
&=&-\frac \delta{\delta\rho}H_c^{(1)}
\non\\
&=&-(\partial_\rho^2V)\varphi_\rho
-\frac 12(\nabla^2\theta)\varphi_\theta
\ee

\no while
\be
G_\theta^{(1)}&=&\{\chi_\theta,H_c^{(1)}\}
\non\\
&=&-\frac \delta{\delta\theta}H_c^{(1)}
\non\\
&=&+\frac 12 (\nabla^2\theta)\varphi_\rho
\ee

\no so that
\be
H_c^{(2)}
=-\frac 12\int dx \;(\partial_\rho^2 V)\varphi_\rho^2.
\ee

\no For third correction to the canonical Hamiltonian, we find
\be
G_\rho^{(2)}&=&\{\chi_\rho,H_c^{(2)}\}
\non\\
&=&-\frac \delta{\delta\rho}H_c^{(2)}
\non\\
&=&\frac 12 (\partial_\rho^3V)\varphi_\rho^2
\ee

\no and 
\be
G_\theta^{(2)}=-\frac \delta{\delta\theta}H_c^{(1)}
=0
\ee

\no so that 
\be
H_c^{(3)}
=\frac 1{2.3}\int dx\; (\partial_\rho^3 V)\varphi_\rho^3.
\ee

\no For the next term we have
\be
G_\rho^{(3)}&=&\{\chi_\rho,H_c^{(3)}\}
\non\\
&=&-\frac \delta{\delta\rho}H_c^{(3)}
\non\\
&=&\frac 1{2.3}(\partial_\rho^4V)\varphi_\rho^3
\ee

\no and $G_\theta^{(n)}=0$ for $n\ge 2$ so that
\be
H_c^{(4)}
=-\frac 1{2.3.4}\int dx\; (\partial_\rho^4 V)\varphi_\rho^4.
\ee


\section*{Appendix B}
\renewcommand{\theequation}{B.\arabic{equation}}
\setcounter{equation}{0}

Let us show here that the first-class constraints $\Omega_\alpha$
are strongly involutive in respect to the canonical Hamiltonian, 
$H_c(\rho,\theta,\varphi_\rho,\varphi_\theta)$, {\sl i. e.},
$\{\Omega_\alpha,H_c\}=0$. 

First note that from 
definition of $\Omega_\rho$, $\Omega_\theta$, Eqs. (\ref{Omegarho}), 
(\ref{Omegatheta}), and the canonical Hamiltonian Eq. (\ref{H-c}), 
we have
\be
\{\Omega_\rho,H_c\}
=&&\{\Omega_\rho,H_c^{(0)}\}
+\{\lambda_\rho,\Omega_\rho\}\varphi_\theta
-\lambda_\theta\{\Omega_\rho,\varphi_\rho\}
\non\\
&&-\{\lambda_\theta,\Omega_\rho\}\varphi_\rho
-\{\Omega_\rho,e^{-\varphi_\rho\partial_\rho}V\},
\ee

\no where we have used the fact that 
$\{\Omega_\rho,\varphi_\theta\}=0$. Then, 
\be
\{\Omega_\rho,H_c\}
=\{\lambda_\rho,\Omega_\rho\}\varphi_\theta
-\{\lambda_\theta,\Omega_\rho\}\varphi_\rho.
\ee

\no Now, using $\varphi_\theta=\Omega_\rho-\chi_\rho$ and 
$\varphi_\rho=\Omega_\theta-\chi_\theta$ we have
\be
\{\Omega_\rho,H_c\}
&=&\{\lambda_\rho,\Omega_\rho\}(\Omega_\rho-\chi_\rho)
-\{\lambda_\theta,\Omega_\rho\}(\Omega_\theta-\chi_\theta)
\non\\
&=&(\{\lambda_\rho,\Omega_\rho\}
         -\{\lambda_\rho,\chi_\rho\})\Omega_\rho
\non\\
&&+(\{\lambda_\theta,\Omega_\theta\}
       -\{\lambda_\theta,\chi_\theta\})\Omega_\rho
\non\\
&=&\{\lambda_\rho,\Omega_\rho\}\Omega_\rho
+\{\lambda_\theta,\Omega_\theta\}\Omega_\rho
\ee

\no and then we find
\be
\{\Omega_\rho,H_c\}=0.
\ee

\no An analogous result can be found for $\Omega_\theta$, proving 
our original statement.



\end{document}